\documentclass[12pt,leqno]{article} 
\usepackage[dvips]{graphics}
\usepackage  {epsfig}      
\newfont{\rams}{msbm10 scaled\magstep1}
\newfont{\fams}{msbm10}
\newfont{\iams}{msbm8}
\newfont{\gotic}{eufm10 scaled\magstep1}
\newfont{\bellap}{eusm10 scaled\magstep1}
\newfont{\bellaps}{eusm7} 
\newcommand{\rea}{\mbox{\rams \symbol{'122}}}

\newcommand{\ecar}{\mbox{\gotic \symbol{'145}}}

\newcommand{\Kb}{\hspace{1pt} \mbox{\bellap K}}

\newcommand{\nq}{{\sf n}_{q}}
\newcommand{\hw}{\hbar \omega}
\newcommand{\kB}{k_{B}}
\newcommand{\TL}{T_{L}}
\newcommand{\kT}{\kB \TL}
\newcommand{\mass}{m^{*}}

\newcommand{\dm}{\displaystyle}
\newcommand{\p}{\hspace{1pt} .}
\newcommand{\sv}{\hspace{1pt} ,}
\newcommand{\ud}{:=}

\newcommand{\bE}{{\bf E}}
\newcommand{\bk}{{\bf k}}
\newcommand{\bka}{{\bf k}_{*}}

\newcommand{\bx}{{\bf x}}

\newcommand{\txk}{(t, \bx, \bk)}
\newcommand{\vk}{{\bf v}(\bk)}
\newcommand{\varf}{\delta f}

\newcommand{\ftxk}{f \txk}
\newcommand{\Ftxk}{F \txk}

\newcommand{\fboo}{g_{0}}
\newcommand{\fboi}{g_{1}}
\newcommand{\fbio}{h_{0}} 
\newcommand{\fbii}{h_{1}} 
\newcommand{\fbooe}{\fboo(\varepsilon)}
\newcommand{\fboie}{\fboi(\varepsilon)}
\newcommand{\fbioe}{\fbio(\varepsilon)}
\newcommand{\fbiie}{\fbii(\varepsilon)}
\newcommand{\dftxk}{\varf \txk}
\newcommand{\hk}{h(\bk)}
\newcommand{\gk}{g(\bk)}

\newcommand{\su}[1]{\sigma(#1)}

\newcommand{\en}{\varepsilon}

\newcommand{\enk}{\en(\bk)}
\newcommand{\enka}{\en(\bka)}

\newcommand{\Rk}{ \rea^{3}}

\newcommand{\devp}[2]{ \frac{\partial #1}{\partial #2}}

\newcommand{\Itre}{\int}

\newcommand{\Ik}[1]{\Itre #1 \hspace{2pt} d \bk}
\newcommand{\Ika}[1]{\Itre #1 \hspace{2pt} d \bka}

\newcommand{\ddm}{\hspace{1pt} \delta (\enka - \enk - \hw) }
\newcommand{\ddp}{\hspace{1pt} \delta (\enka - \enk + \hw) }
\newcommand{\dee}{\hspace{1pt} \delta (\enka - \enk) } 
\newcommand{\dpm}{\hspace{1pt} \delta (\enka - \enk \pm \hw) }
\newcommand{\deu}{\hspace{1pt} \delta (\enk - u) }

\newcommand{\eq}[1]{\mbox{{\rm(\ref{#1})}}}
\newcounter{nsez}
\newcommand{\sez}[1]{\\[10pt] {\large \bf \addtocounter{nsez}{1}
             \noindent \thensez. #1} \\}
\newcounter{nsbs}

\newcommand{\cita}[1]{\cite{#1}} 
\newcommand{\auto}[2]{{\sc #2} {\sc #1} }
\newcommand{\arti}[5]{\hspace{-6pt}, {\it #1}, {#2} {#5}; {#3}: #4.}
\newcommand{\libr}[4]{\hspace{-6pt}, {\it #1}, #2, #3, #4. }
\begin{document}
\baselineskip=8mm
\jot=8pt
\arraycolsep=1pt
\begin{center}
{\large \bf High Field Mobility and Diffusivity of 
Electron Gas in Silicon Devices}
\\[3mm]
{\sc S.~F.~Liotta\footnote{e-mail: liotta@dipmat.unict.it} 
and A.~Majorana\footnote{e-mail: majorana@dipmat.unict.it}} 
\\
{Dipartimento di Matematica - University of Catania - } \\
{ Viale A.~Doria 6, 95125 Italy}
\end{center}
{\bf Abstract.}
In this paper the Boltzmann equation describing the carrier
transport in a semiconductor is considered. 
A modified Chapman-Enskog method is used, in order to find approximate
solutions in the weakly non-homogeneous case. 
These solutions allow to calculate the mobility and 
diffusion coefficients as function of the electric field.
The integral-differential equations derived by the above method are
numerically solved by means of a combination of spherical harmonics
functions and finite-difference operators.
The Kane model for the electron band structure is assumed;
the parabolic band approximation is obtained
as a particular case. 
The numerical values for mobility and diffusivity in a
silicon device are compared with the experimental data.
The Einstein relation is also shown.\\[15pt]
\sez{Introduction}
Many commercial simulators of microelectronic devices are based on
the well-known Drift-Diffusion equations.
In this model the electron current density ${\bf J}_{n}$ 
is given by the equation~\cita{Ferr}, \cita{Hans} 
\begin{equation}
   {\bf J}_{n} = \mu_{n} \rho \bE + D_{n} \nabla_{\bx}\rho  \sv
   \label{e1}
\end{equation}
where $\rho$ is the electron density, $\bE$ the electric field, 
$\mu_{n}$ and $D_{n}$ the mobility and the diffusion coefficient,
or diffusivity, respectively. The symbol  
$\nabla_{\bx}$ denotes the gradient operator
with respect to ${\bf x}$.
For low electric field the two transport coefficients $\mu_{n}$ and  
$D_{n}$ are related by the well-known Einstein relation 
\begin{equation}
 D_{n} =\left(\frac{\kT}{\ecar}\right)\mu_{n} 
 \label{e2}
\end{equation}
where $\kB$ is the Boltzmann constant, $\TL$ the lattice temperature and
$\ecar$ the absolute value of the electron charge.
The transport coefficients $\mu_{n}$ and $D_{n}$ are often 
assumed constant. This approximation becomes inadequate in presence of
high electric fields. In these cases Eq.~\eq{e1} remains valid only if 
the transport coefficients are considered as functions of the electric
field. Many expression for these functions were proposed (see, for instance, Refs.
\cita{Hans}, \cita{MaRi}, \cita{Selb}).
They were obtained by fitting experimental data or Monte Carlo
simulations \cita{JaLu},~\cita{Tomi}. 
In this paper we obtain the functions $\mu_{n}(E)$ and $D_{n}(E)$ 
directly from the Boltzmann transport equation.\\
 The paper is organised as follows. In Sec.~2 we briefly introduce the
Boltzmann transport equation for an electron gas in a semiconductor. 
Only electron-phonon scatterings are considered. So that, electron-electron
and electron-impurity interactions are assumed negligible.
In Sec.~3 we perform a Chapman-Enskog expansion, 
and in Sec.~4 a numerical scheme is proposed in order to
solve the obtained equations.
In the last section we show numerical results
for $\mu_{n}$ and $D_{n}$, which we  
compare with experimental data. 
Further, we analyze the validity of the
Einstein relation for different values of the electric field. 
\sez{Basic equations} 
For an electron gas in a semiconductor device the 
coupled system Boltzmann-Poisson equations  
\cita{Ferr} writes
\begin{eqnarray}
 \devp{F}{t} +\vk \cdot \nabla_{\bx} F -
 \frac{\ecar}{\hbar} \bE \cdot \nabla_{\bk} F = Q(F) ,
 \label{e3} \\
 \nabla_{\bx} \cdot \bE =\frac{\ecar}{\epsilon}
 \left(N_{D}(\bx)-N_{A}(\bx) - {\Ik F}\right)
 \label{e3bis}
\end{eqnarray}
where the unknown function $\Ftxk $ 
represents the probability of finding an electron at the position 
$\bx$, with wave-vector $\bk$, at time $t$.
The wave-vector $\bk$ belongs to $\Rk$.
The integrals with respect to $\bk$ are performed over 
the whole space and the parameter $\hbar$ is the Planck 
constant divided by $2\pi$.
The group velocity $\vk$ depends on the band structure
and it will be defined in the following.
The symbol  $\nabla_{\bk}$ denotes the gradient with 
respect to the variables $\bk$. 
The functions $N_{A}$~,~$N_{D}$ are the concentration 
of acceptors and donors and  
$\epsilon$ is the dielectric constant.
Here we assume that the low-density approximation
holds, so that $Q$ is a linear operator. If $K(\bk,\bka)$ is the 
symmetric scattering kernel, then
\begin{eqnarray}
Q(F) \txk =  \Ika{K(\bk,\bka) F(t,\bx,\bka)} 
 - \Ftxk \Ika{K(\bka,\bk)}  \label{e3tris}   \p
\end{eqnarray}
We consider the scattering kernels describing
acoustic phonon interactions (in the elastic approximation)
\begin{displaymath}
K_{ac}(\bka,{\bf k}) = {\cal {G}}~\Kb_{0} \dee \sv
\end{displaymath} 
and non-polar optical phonon interactions 
\begin{displaymath}
K_{op}(\bka,{\bf k}) = \left( \begin{array}{c} \nq +1 \\ \nq 
\end{array} \right) {\cal {G}}~\Kb \dpm \p
\end{displaymath}
Here, $\Kb_{0}$ and $\Kb$ are constant and
the overlap factor ${\cal {G}}$ for the conduction band is taken equal
to $1$ (see~\cita{JaLu},~\cita{ReVe}).
We consider negligible the ionized impurity scattering; i.e. we assume
low density of doping.
With these assumptions,
the collision operator esplicitelly writes 
\begin{eqnarray}
& & 
 Q(f) \txk = (\nq +1) \Ika{\Kb \ddm  f(t, \bx, \bka) } 
\nonumber \\
& & 
\mbox{ } + \nq
\Ika{ \Kb \ddp  f(t, \bx, \bka) } 
\nonumber \\
& & 
\mbox{ } + \Ika{ \Kb_{0} \dee f(t, \bx, \bka) } - \bar{\nu}(\bk) \ftxk \sv
\label{e50}  
\end{eqnarray}
where
\begin{eqnarray}
& & \mbox{} \hspace{-5pt} \bar{\nu}(\bk) = \nq \Ika{\Kb \ddm} + (\nq + 1) 
\nonumber \\
& & \mbox{} \times \Ika{\Kb \ddp} + \Ika{\Kb_{0} \dee} \p 
 \label{e51} 
\end{eqnarray}
For every admissible function $\Ftxk$, we have~\cita{Maj1}
\begin{displaymath}
\Ik{Q(F)} = 0  \sv
\end{displaymath}
which implies mass conservation.
The band structure is assumed to be spherically symmetric. 
Hence, the electron energy $\en$ is given by the equation 
\begin{equation}   
\gamma(\en) = \frac{\hbar^{2}}{2 \mass} k^{2} \sv
\label{e5}
\end{equation}
where $\mass$ is the reduced electron mass. 
We choose the Kane formula
\begin{equation}   
\gamma(\en) = \en (1 + \alpha \en)  \sv
\label{e6}
\end{equation}
which is simple but sufficiently accurate to describe high field regime.
By putting $\alpha=0$ one obtain the usual parabolic band approximation. 
The velocity $\vk$ is explicity given by
\begin{equation}
\vk \ud \frac{1}{\hbar} \nabla_{\bk} \enk 
= \frac{\hbar \bk}{\mass (2\alpha \en +1)} 
\label{e6bis} \p
\end{equation}
\sez{The Chapman-Enskog expansion}
The Chapman-Enskog method \cita{LaLi} allows us to find 
approximate solutions of the Boltzmann equation. They are 
valid only for a very small space domains. 
Moreover, these regions must be far enough
from boundary layers. Nevertheless, these solutions have a 
great importance since they allow to obtain the transport coefficients. 
In the framework of electron transport in semiconductors, 
we expect that such solutions are valid in spatial domain, where the
electric field and the doping are almost constant. 
Therefore, they should not furnish good 
results near junctions or boundaries.\\  
We follow the standard scheme \cita{LaLi}, assuming that the 
solution of Eq.~\eq{e3} is approximately expressible by the following
relation
\begin{equation}
\Ftxk \simeq \ftxk + \dftxk \p \label{e7}
\end{equation}
As usually, the typical Chapman-Enskog constrain
\begin{equation}
\Ik \dftxk = 0  \label{e8} 
\end{equation}
is imposed. Therefore
using Eqs.~\eq{e7}-\eq{e8} we obtain that the electron density is
\begin{displaymath}
\rho(t,\bx) \ud \Ik{\Ftxk} \simeq \Ik{\ftxk} \p
\end{displaymath}
We assume that {\em time and space 
partial derivatives of $f$ are of the same order of $\varf$} and
that the {\em corresponding derivatives of $\varf$ are negligible}.
By inserting Eq.~\eq{e7} into   
Eq.~\eq{e3} and splitting terms of
different orders, we obtain the
following equations 
\begin{eqnarray}
 & &   -\frac{\ecar}{\hbar} \bE \cdot \nabla_{\bk} f = Q(f) \sv
\label{e9} \\
 & & \devp{f}{t} +\vk \cdot \nabla_{\bx} f -
 \frac{\ecar}{\hbar} \bE \cdot \nabla_{\bk} \varf = Q(\varf) 
 \label{e10}  \p
 \end{eqnarray}
In this approximation,  the Poisson equation becomes   
\begin{equation}
\nabla_{\bx} \cdot \bE \simeq \frac{\ecar}{\epsilon}
\left(N_{D}-N_{A} - {\Ik {f}}\right)
\label{e11}  \p
\end{equation}
In order to solve Eqs.~\eq{e9}-\eq{e10} we assume that $\bE$ is
constant. This is a reasonable assumption, if the difference 
between the density of electrons and 
doping is negligible in the small
domains, where we consider the Boltzmann equation. \\
Of course, we solve before Eq.~\eq{e9} and then the next one.
Since Eq.~\eq{e9} is linear and it does not depend 
explicitly on the variables $(t,\bx)$,
a solution can be written as
\begin{equation}
\ftxk = \rho(t,\bx) \gk \label{e12} \p
\end{equation}
We note that  the function $\gk$ is also 
the stationary homogeneous solution of 
Eq.~\eq{e3} for constant electric field, verifying the condition 
\begin{equation}
{\Ik {\gk}} = 1 \label{e13} \p
\end{equation}
Now, we consider Eq.~\eq{e10}. Using~\eq{e12}, it becomes 
\begin{equation}
 \gk 
 \left[ \devp{\rho}{t} +\vk \cdot \nabla_{\bx}\rho \right] 
- \frac{\ecar}{\hbar} \bE \cdot \nabla_{\bk} \varf = Q(\varf) 
 \label{e15}  \p
 \end{equation}
By integrating Eq.~\eq{e15} with respect to $\bk$, we find a
compatibility condition, as usually arises in integral equations.
Assuming the reasonable hypotesis 
$\int \nabla_{\bk} \varf d \bk = 0$, we obtain  
\begin{equation}
\devp{\rho}{t} +{\bf V} \cdot \nabla_{\bx}\rho = 0 \sv
\label{e16}
\end{equation}
where
\begin{equation}
{\bf V} \ud  {\Ik {\gk \vk} } \label{e17}
\end{equation}
is the constant macroscopic electron velocity 
in the stationary homogeneous case. \\
This vector is parallel to $\bE$ (see appendix A).
By eliminating the time derivative of $\rho$ in 
Eq.~\eq{e15} by using Eq.~\eq{e16}, we get
\begin{equation}
 \gk 
 \left[ \vk - {\bf V}  \right] \cdot \nabla_{\bx}\rho 
- \frac{\ecar}{\hbar} \bE \cdot \nabla_{\bk} \varf = Q(\varf) 
 \label{e18}  \p
\end{equation}
The form of Eq.~\eq{e18} suggest us to assume  
\begin{equation}
\varf(t,\bx,\bk) =|\nabla_{\bx}\rho(t,\bx)| \hk  \label{e19} \p
\end{equation}
Now, Eq.~\eq{e8} implies that 
\begin{equation}
\Ik \hk = 0 \label{e20} \p
\end{equation}
We denote by  ${\bf u}$ the unit vector in the 
direction of $\nabla_{\bx}\rho$; i.e. 
$\nabla_{\bx}\rho = |\nabla_{\bx}\rho| {\bf u}$.
It in general may depend on the variables $(t,\bx)$. 
Using Eq.~\eq{e19}, it is easy to see that Eq.~\eq{e18} becomes 
\begin{equation}
 \gk 
 \left[ \vk \cdot {\bf u} - {\bf V} \cdot {\bf u} \right] 
- \frac{\ecar}{\hbar} \bE \cdot \nabla_{\bk} h = Q(h) 
\label{e21} \p
\end{equation}
The case ${\bf u}$ parallel to $\bE$ is the most meaningfull.
\sez{Approximate equations}
In order to obtain  $f$ and $\varf$ we have to
solve the following set of equations for the unknowns $g$ and $h$ 
\begin{eqnarray}
 & &-\frac{\ecar}{\hbar} \bE \cdot \nabla_{\bk} g = Q(g) \sv
\label{e22} \\
& & \gk  \left[ \left| \vk \right| \cos \theta - V  \right]   
- \frac{\ecar}{\hbar} \bE \cdot \nabla_{\bk} h = Q(h) 
 \label{e23}  \sv  \\
 & &{\Ik {\gk}} = 1  ~~~,~~~ \Ik \hk = 0 \p \label{e223}
\end{eqnarray}
Here $\theta$ is the angle between $\vk$ and ${\bf u}$, and
$V = |{\bf V}|$.\\
Since $g$ and $h$ are functions of the 3-dimensional variable $\bk$, we use
a spherical harmonics expansion to reduce the dimension of the space of the 
indipendent variables.
Taking in account the symmetries of the problem and considering only the first
two terms of the expansion, we can choose 
\begin{eqnarray}
\gk \simeq  \fbooe + \fboie \cos \theta \label{e24} \sv \\
\hk \simeq  \fbioe + \fbiie \cos \theta \label{e25} \p
\end{eqnarray}
The use of the Galerkin method allows to derive 
from Eqs.~\eq{e22}-\eq{e23} a set of equations
for $\fboo, \fboi, \fbio$ and $\fbii$. This approach
recalls the well-known method of the spherical harmonics 
expansion to solve the Boltzmann-Poisson 
system,  \cita{LiMa1},  \cita{An1},  \cita{VeBa}. 
Since Eqs.~\eq{e22}-\eq{e23} contain the same diffusion and
collisions terms of the Boltzmann Equation, many calculations
are performed following Ref.~\cita{An1}.
Then, the system~\eq{e22}-\eq{e23} gives 
the following set of ordinary differential-difference equations
\begin{eqnarray}
& & -\ecar E 
\left(\frac{\gamma'(\en)}{\gamma(\en)}\fboi + \fboi' \right) 
 \frac{3}{v(\en)} = Q_{1}(\fboo) \label{e26} \\
& &-\ecar E \fboo'  \frac{1}{v(\en)}=Q_{2}(\fboi) \label{e27} \\
& & -V \fbooe \frac{3}{v(\en)} + \fboie
-\ecar E 
\left(\frac{\gamma'(\en)}{\gamma(\en)}\fbii + \fbii' \right) 
 \frac{3}{v(\en)}=Q_{1}(\fbio) \label{e28} \\
& & -V \fboie \frac{1}{v(\en)} + \fbooe
-\ecar E \fbio' =  \frac{1}{v(\en)}=Q_{2}(\fbii) \label{e29}
\sv
\end{eqnarray}
where the prime indicates the derivative 
with respect to $\en$, 
\begin{displaymath}
 v(\en) \ud \sqrt{\frac{2}{\mass}} 
\frac{\sqrt{\gamma(\en)}}{\gamma'(\en)} \sv 
\end{displaymath}
and 
\begin{eqnarray}
Q_{1}(\varphi) &=& (\nq + 1) \Kb~\su{\en+\hw} \varphi(\en+\hw) 
+ \nq \Kb~\su{\en-\hw} \varphi(\en-\hw)  
\nonumber \\
&{}& - \left[ \nq \Kb~\su{\en+\hw} + (\nq + 1) \Kb~\su{\en-\hw} 
\right] \varphi(\en) \label{e30} \sv \\
Q_{2}(\varphi) &=&   
-\left[\nq \Kb~\su{\en+\hw} \right. + (\nq + 1) \Kb~\su{\en-\hw} + 
\left. \Kb_{0}~\su{\en} \right]\varphi(\en)  
\label{e31} \p  
\end{eqnarray}
The density of states $\su{u}$ is given by
\begin{displaymath}
\su{u} \ud \Ik{\deu} =  \nonumber \\
4 \sqrt{2} \pi \left( \frac{\sqrt{\mass}}{\hbar} 
\right)^{3} H(u) \sqrt{\gamma(u)} 
\gamma'(u)  \sv
\end{displaymath}
where $H(u)$ is the Heaviside step function. 
\sez{Physical model and numerical results} 
We use the following values for the parameters, appropriate for a
silicon device:
\begin{center}
\begin{tabular}{|l|l|l|}
\hline
$ \mass = 0.32 \, m_{e}$ & $ \TL = 300 $ K & $\hw = 0.063$ eV \\[7pt]
$ \dm \Kb = \frac{\left( D_{t} K \right)^{2}}{8 \pi^{2} \rho \omega} $ &
$ D_{t} K = 11.4 $ eV \mbox{$\stackrel{\circ}{\mbox{\rm A}}$}$^{-1}$ &
$\rho = 2330$ Kg m$^{-3}$ \\[15pt]
$ \dm \Kb_{0} = \frac{\kT}{4 \pi^{2} \hbar v_{0}^{2} \rho} \Xi_{d}^{2} $ 
& $\Xi_{d} = 9$ eV & $v_{0} = 9040$ m sec$^{-1}$. \\[15pt]
$ \alpha = 0.5 \, eV^{-1}$ & $ \mbox{ } $ & $ \mbox{ } $  \\[7pt]
\hline
\end{tabular}
\end{center}
In the table, $m_{e}$ is the electron rest mass. 
We have solved numerically Eqs.~\eq{e26}-\eq{e29} 
with the conditions \eq{e223} for
different values of the electric field.
The numerical procedures are similar to that 
used in Ref.~\cita{LiMa1}. The kinetic definition of 
the current density is  
\begin{equation}
{\bf J}_{n} \ud  \Ik {F(t,\bx,\bk) \vk} \simeq \rho(t,\bx) {\bf V} 
+| \nabla_{\bx}\rho (t,\bx) | {\Ik {\hk \vk} }  \p \label{e32}
\end{equation}
Since ${\bf V} = V {\bf u}$ and
\begin{eqnarray}
{\Ik {\hk \vk} } \simeq { \Ik {[\fbioe + \fbiie \cos \theta] \vk }} \nonumber \\
 = {\Ik { \fbiie \frac{\vk \cdot {\bf u}}{|\vk|} \vk}} 
 =  {\bf u}{\Ik { \fbiie \frac{(\vk \cdot {\bf u})^{2}}{|\vk|}}}  \sv \label{e33}
\end{eqnarray}
a comparison between Eqs.~\eq{e32}-\eq{e33} and \eq{e1}
gives
\begin{equation}
\mu_{n} = \frac{V}{|\bE|} ~~~and~~~ D_{n} = 
{\Ik { \fbiie \frac{(\vk \cdot {\bf u})^{2}}{|\vk|}}} \label{e34} \p
\end {equation}
They are functions of $E$, because
$V$ and $\fbii$ depend on this parameter.\\
In fig.~1 we show the mobility $\mu_{n}$ against the electric field.
The values are compared with two curves which fit experimental data
by means of simply formulas (see~\cita{Selb}, pp.~94-98).
For low values of the electric field there is a small difference. It is
meaningless because many different values (up to 20 {\%} ) of low field
mobility are reported in literature (see~\cita{Selb}, pp.~81-82). 
In fig.~2 the value of diffusivity is shown together with experimental data
(see~\cita{BJNR}, p.~6715). The agreement is very good for all range of $|\bE|$.
Fig.~3 and fig.~4 compare the mobility and the diffusivity 
using Kane model and the parabolic
band approximation. The last figure shows that Einstein 
relation is valid only for moderate fields. \\[25pt]
{{\large \bf Acknowledgments}
\\
We acknowledge partial support from Italian Consiglio Nazionale
delle Ricerche (Prog. N. 97.04709.PS01) and TMR project 
n.~ERBFMRCT970157 ``Asymptotic Methods in Kinetic Theory''.
\\[25pt]
\large \bf Appendix A} 
\\
The unknown $g$ must satisfy Eq.~\eq{e22}, which contains only two
vectors, $\bk$ and $\bE$. Since g is a scalar function, the only
possibility is that $g$ depends on $\bk$ only throught the scalars
$|\bk|$ and $\bk \cdot \bE$ i.e. $g=\tilde{g}(|\bk|, \bk \cdot \bE)$.
Therefore, from Eq.~\eq{e17} we have
\begin{displaymath}
{\bf V} \ud  {\Ik {\gk \vk}} 
= {\Ik {\tilde{g}(|\bk|, \bk \cdot \bE)
\frac{\hbar \bk}{\mass (2\alpha \en +1)}}} \p
\end{displaymath}
Hence, it is evident that ${\bf V}$ is parallel to $\bE$.\\
%

%
%
%
%
%
%

%
\newpage
\begin{figure}[h!]
\centerline{\psfig{figure=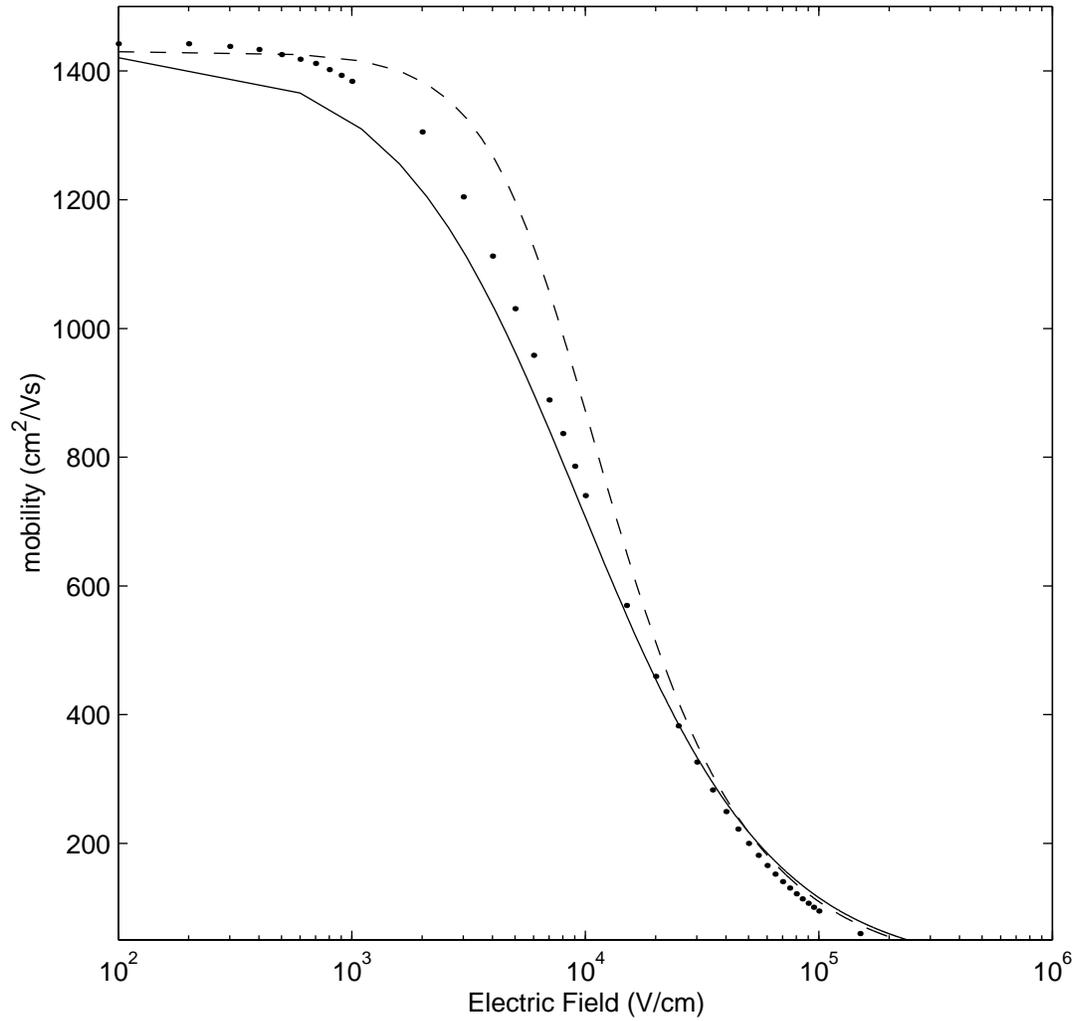}}
\caption{ \small Mobility as a function of electric field: dots indicates values
obtained using Kane band model,  solid line a fit with parameters
of Canali et. al. and dashed line another fit with parameters of
Caughey and Thomas (see Ref.~\cita{Selb}).}
\end{figure}
\newpage
\begin{figure}[h!]
\centerline{\psfig{figure=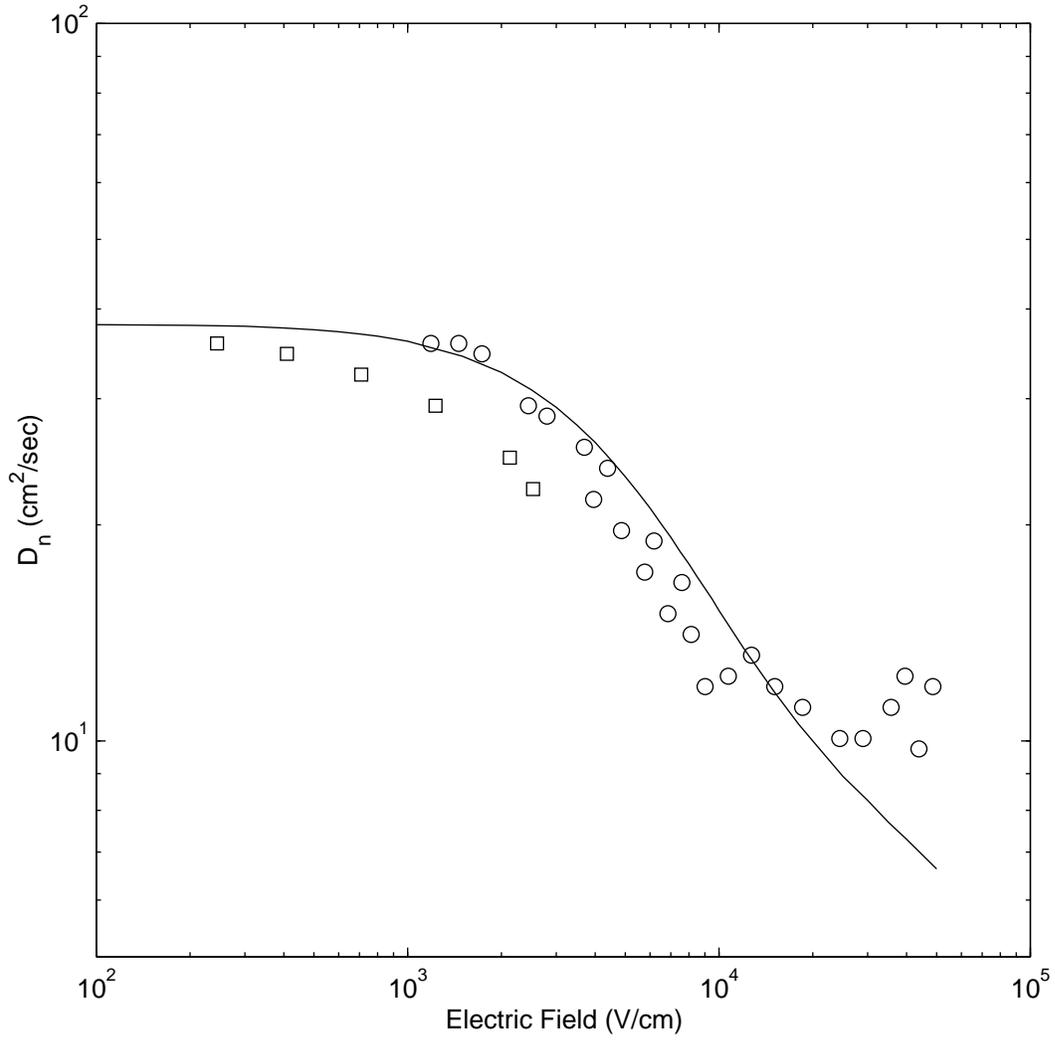}}
\caption{\small Diffusivity as a function of electric field: solid line indicates  
values calculated using Kane band model and squares and circles 
experimental data (see Ref.~\cita{BJNR}).}
\end{figure}
\newpage
\begin{figure}[h!]
\centerline{\psfig{figure=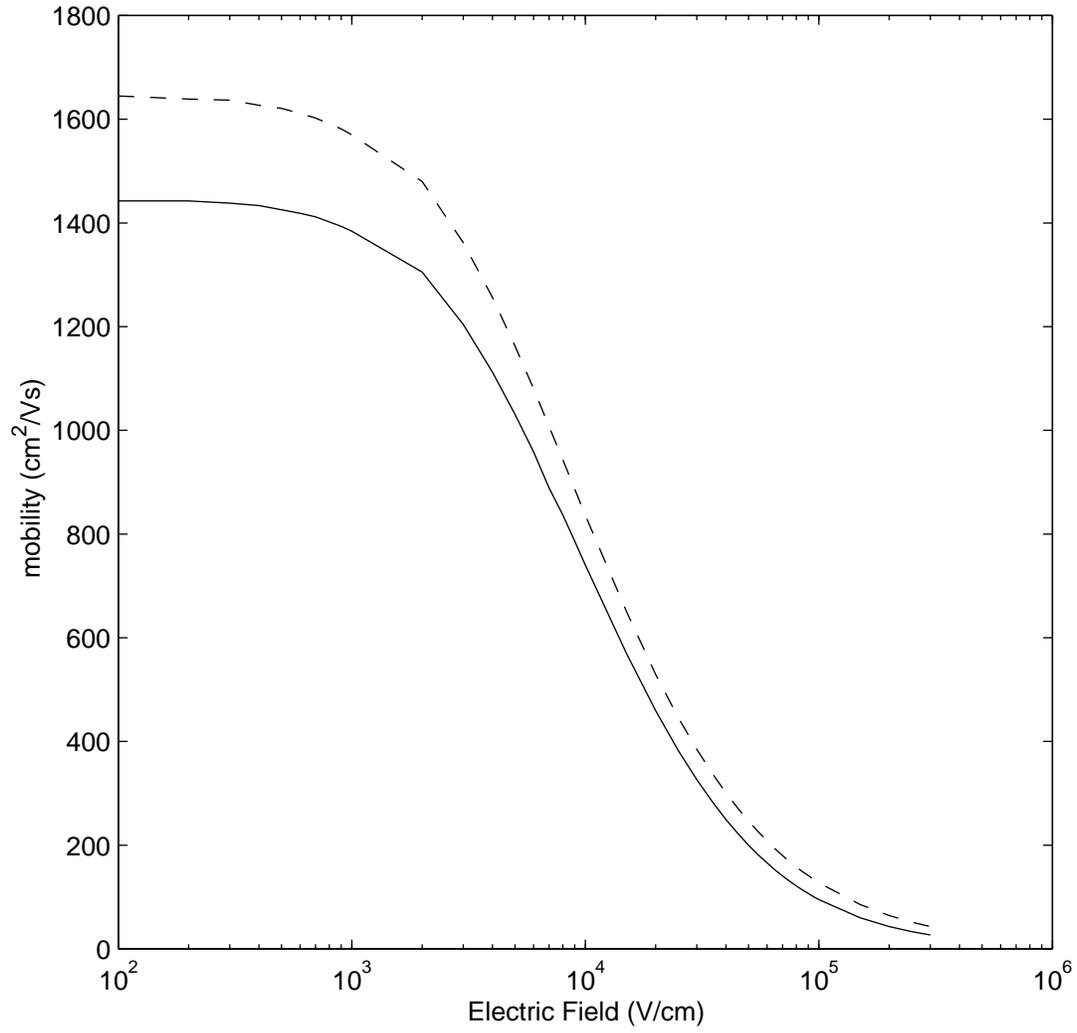}}
\caption{\small Mobility versus electric field: 
Kane band model (solid line) and parabolic band approximation (dashed 
line).}
\end{figure}
\newpage
\begin{figure}[h!]
\centerline{\psfig{figure=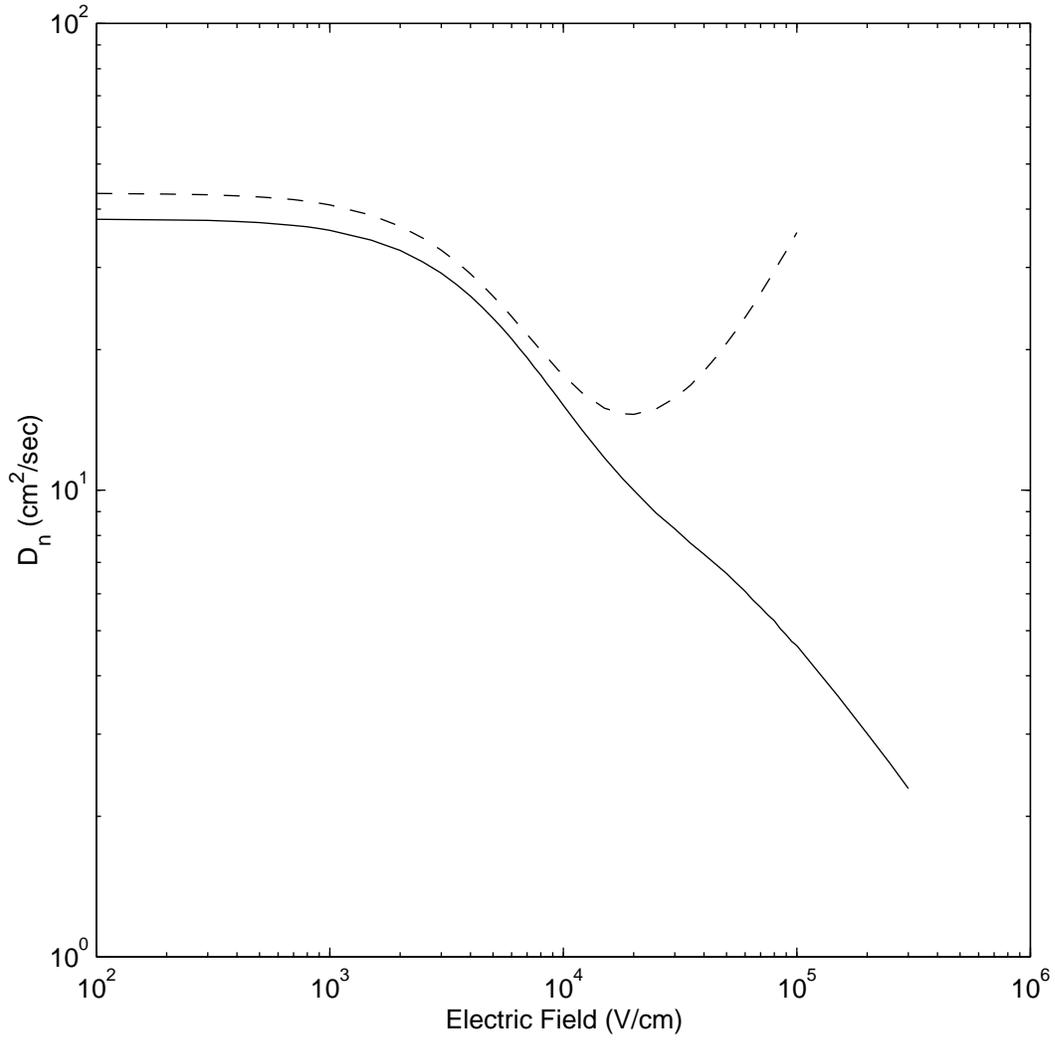}}
\caption{\small Diffusivity versus electric field: 
Kane band model (solid line) and parabolic band approximation (dashed 
line).}
\end{figure}
\newpage
\begin{figure}[h!]
\centerline{\psfig{figure=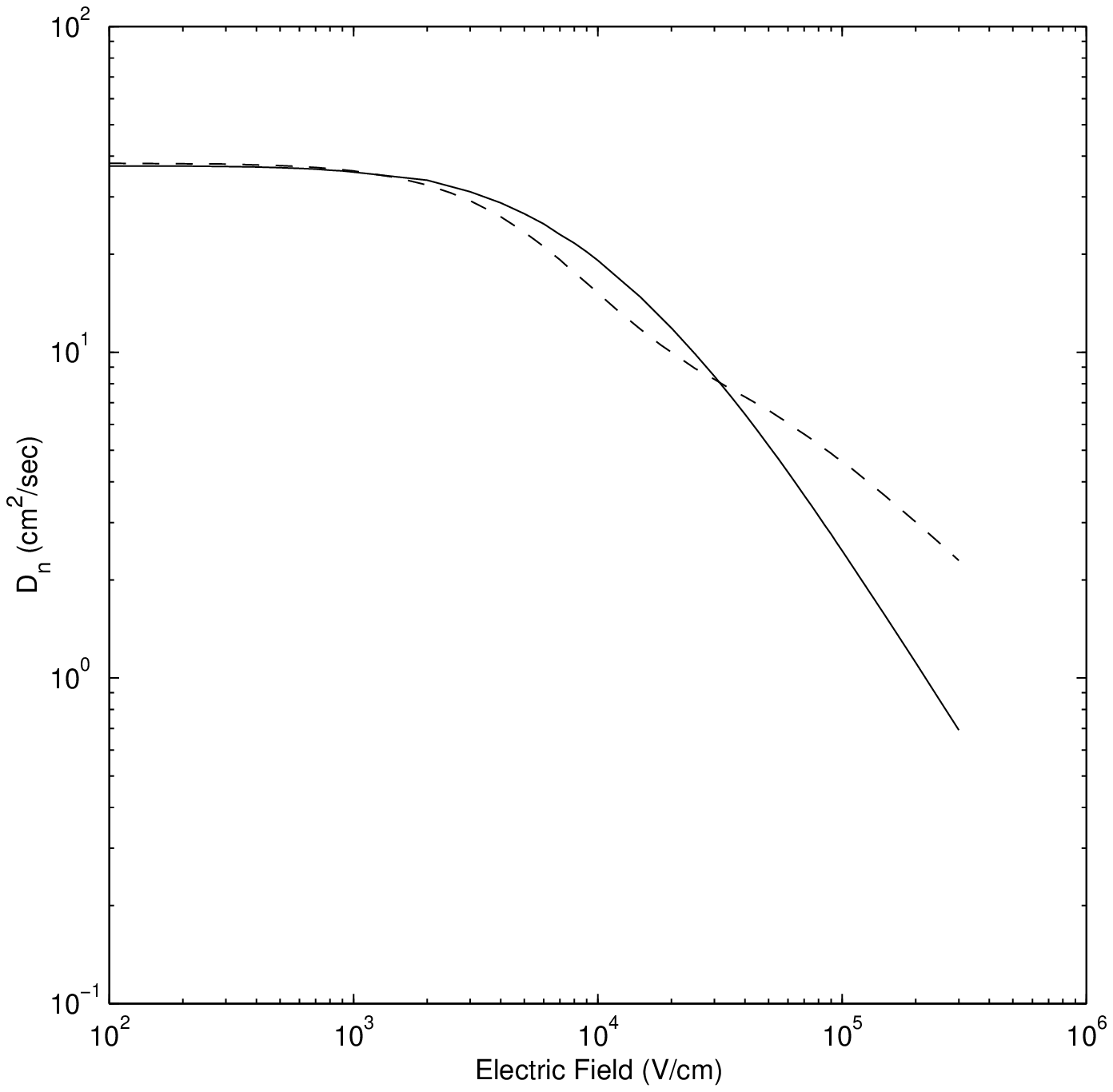}}
\caption{\small Comparison between diffusivity calculated directly (dashed line)
and using Einstein relation (solid line), 
both with Kane band model.}
\end{figure}


\newpage
\small
\begin{thebibliography}{33}
%
\bibitem{BJNR} 
\auto{R.}{Brunetti}, \auto{C.}{Jacoboni}, \auto{F.}{Nava}, 
\auto{L.}{reggiani}, \auto{G.}{Bosman} and \auto{R.J.J.}{Zijlstra}
\arti{Diffusion Coefficient of Electrons in Silicon}
{J. Appl. Phys}{52}{6713-22}{1981}
%
\bibitem{Ferr} 
\auto{D.~K.}{Ferry}
\libr{Semiconductors}{New York}{Macmillan Publ. Comp.}{1991} 
%
\bibitem{Hans}
\auto{W.}{H\"{a}nsch} 
\libr{The Drift-Diffusion Equation and its Applications in MOSFET 
Modelling}{New York}{Springer-Verlag}{1991} 
%
\bibitem{JaLu}
\auto{C.}{Jacoboni} and \auto{P.}{Lugli} 
\libr{The Monte Carlo Method for Semiconductor Device Simulation}{New 
York}{Springer-Verlag}{1989}
%
\bibitem{LaLi}
\auto{E.M.}{Lifshitz} and \auto{L.P.}{Pitaevskij} 
\libr{Physical Kinetics, vol. 10, in the Landau \& Lifshitz
course of theoretical physics}{Oxford} 
{Pergamon press}{1981}
%
\bibitem{LiMa1} 
\auto{S.~F.}{Liotta} and \auto{A.}{Majorana} 
\libr{A Novel Approach to Spherical-Harmonics Expansion for Electron Transport in
Semiconductors}{}{Preprint
Univ. of Catania}{1999}
%
\bibitem{Maj1} 
\auto{A.}{Majorana} 
\arti{Trend to Equilibrium of Electron Gas in a Semiconductor
According to the Boltzmann Equation}{Transport Theory 
Statist. Phys.}{27}{547-71}{1998}
%
\bibitem{MaRi}
\auto{P.~A.}{Markowich}, \auto{C.}{Ringhofer} and \auto{C.}{Schmeiser} 
\libr{Semiconductor Equations}{New York}{Springer-Verlag}{1990}
%
\bibitem{An1}
\auto{K.}{Rahmat}, \auto{J.}{White} and \auto{D.~A.}{Antoniadis} 
\arti{Simulation of Semiconductor Devices Using a Galerkin/Spherical 
Harmonic Expansion Approach to Solving the Coupled Poisson-Boltzmann
System}{IEEE Trans. Computer-Aided Design}  
{15}{1181-96}{1996}
%
\bibitem{ReVe} 
\auto{S.}{Reggiani}, \auto{M.~C.}{Vecchi} and \auto{M.}{Rudan}
\arti{Investigation on Electron and Hole Transport 
Properties Using the Full-Band Spherical-Harmonics
Expansion Method}{IEEE Trans. 
on Electron Devices}{45}{2010-17}{1998}
%
\bibitem{Selb}
\auto{S.}{Selberherr} 
\libr{Analysis and Simulation of Semiconductors Devices}
{Wien-New York}{Springer-Verlag}{1984}
%
\bibitem{Tomi}
\auto{K.}{Tomizawa}  
\libr{Numerical Simulation of Sub Micron Semiconductor Device}{Boston} 
{Artech House}{1993}
%
\bibitem{VeBa} 
\auto{D.}{Ventura}, \auto{A.}{Gnudi} and \auto{G.}{Baccarani}
\arti{A Deterministic Approach to Solution of the BTE in
Semiconductors}{Rivista del Nuovo Cimento}{18}{1-33}{1995}
%
\end{thebibliography}
\end{document}